\documentclass[runningheads,a4paper]{llncs}
\usepackage{makeidx}  
\usepackage[misc,geometry]{ifsym}  
\usepackage{graphicx} 
\usepackage{booktabs}
\usepackage{multirow}
\usepackage{siunitx}
\usepackage{cite}
\begin{document}
\frontmatter          
\pagestyle{headings}  
\addtocmark{Hamiltonian Mechanics} 

\mainmatter              
\title{Kidney tumor segmentation using an ensembling multi-stage deep learning approach. A contribution to the KiTS19 challenge}
\titlerunning{Deep Learning for kidney tumor segmentation}  
%
\author{Gianmarco Santini\inst{1}, No\'emie Moreau\inst{1} and Mathieu Rubeaux\inst{1}}
%
\authorrunning{G.Santini et al.}
%
%
\institute{Keosys Medical Imaging, Nantes, France}

\maketitle              
\bigskip
\begin{abstract}
Precise characterization of the kidney and kidney tumor characteristics is of outmost importance in the context of kidney cancer treatment, especially for nephron sparing surgery which requires a precise localization of the tissues to be removed. The need for accurate and automatic delineation tools is at the origin of the KiTS19 challenge. It aims at accelerating the research and development in this field to aid prognosis and treatment planning by providing a characterized dataset of 300 CT scans to be segmented. To address the challenge, we proposed an automatic, multi-stage, 2.5D deep learning-based segmentation approach based on Residual UNet framework. An ensambling operation is added at the end to combine prediction results from previous stages reducing the variance between single models. Our neural network segmentation algorithm reaches a mean Dice score of 0.96 and 0.74 for kidney and kidney tumors, respectively on 90 unseen test cases. 
The results obtained are promising and could be improved by incorporating prior knowledge about the benign cysts that regularly lower the tumor segmentation results.
\medskip
\end{abstract}
\bigskip
\section{Introduction}
Kidney cancer represents 2.4\% of the cancers worldwide with more than 400 000 new cases in 2018 \cite{bray2018global}, with large variations in incidence rates based on geography, ethnicity, gender and over the time \cite{scelo2018epidemiology}. In the last two decades, the growing use of medical imaging had two effects: the renal tumor size diagnosed has consistently decreased \cite{nguyen2006evolving}, and the number of localized renal masses found incidentally, on the other hand, has raised \cite{sun2012treatment}.

While the standard for renal cancer treatment has been traditionally radical nephrectomy (i.e. removal of both the tumor and damaged kidney), more preservative nephronsparing surgery (i.e. open or laparoscopic partial nephrectomy) has more than quadrupled in the last 20 years \cite{nguyen2006evolving} and is now a treatment of choice \cite{dominguez2011laparoscopic}.

In this context, the information related to the kidney and tumor positions, shapes and sizes, is of outmost importance for the surgery evaluation and planning process. While imaging modalities such as CT allow precise detection of the tumors, there is still a lack of automatic delineation tools, and the evaluation continues to rely on the use of nephrometry scoring systems based on manual and relatively simple imaging features \cite{ficarra2009preoperative, kutikov2009renal, simmons2010kidney}. 

A series of methods have been proposed in literature to perform automatic kidney segmentation in various imaging modalities. When dealing with CT, traditional methods based on imaging features, deformable models, active shape models as well as atlases have been suggested \cite{torres2018kidney}. 
More recently the global interest towards deep learning algorithms has led to an incredible variety of applications in the medical imaging field \cite{litjens2017survey}
, and the area of kidney segmentation is no exception \cite{thong2018convolutional, zheng2017deep, jackson2018deep, sharma2017automatic }. 

However, most of these achievements mainly focus on the automatic kidney segmentation, generally on healthy patients, without considering the specificity of cancerous tissue if present. Indeed, the renal tumor characterization can be challenging because of its variety in terms of position, extension and gray scale values. Moreover, its distinction with benign renal cysts is not always trivial, as illustrated by the IIF (Follow-up) categorization of the reference Bosniak classification \cite{bosniak1986current}, where the cysts have a 5-10\% risk of being a kidney cancer.

A first attempt to identify and segment kidney cancer was proposed by \cite{kim2004computer}
, consisting in a computer-aided method able to distinguish healthy renal tissue from cancerous one in CT scans, by using a region growing algorithm and a gray level thresholding.

More recently, Zhou et al. \cite{zhou2016atlas} 
proposed another approach where a two-step segmentation strategy was developed by employing a single atlas model to isolate kidneys as first action, and then using supervoxels to identify possible cancerous areas from the previous segmentation. However, both methods require user interactions; in the first case to initialize some seeds for the region growing, whilst in the second work the user is demanded to manually select suspect regions from the super-voxel probability maps. Moreover in \cite{zhou2016atlas}, only cases where an evident contrastographic difference between kidneys and tumor existed were considered and succeeded in the final segmentation.

In this context, the KiTS19 challenge \cite{heller2019kits19} 
was proposed to stimulate the research and development of trustworthy kidney and kidney tumors segmentation methods, by providing a database of 300 annotated CT scans.

In this paper we propose an automatic segmentation method based on a multi-stage 2.5D deep learning approach to address the KiTS19 MICCAI challenge on tumor kidney segmentation. The rest of the paper is organized as follows. Section 2 presents a detailed overview of the data and methods employed. Section 3 describes the configuration used for training our model and gives the results obtained. Section 4 opens a discussion on the results and presents the conclusions.

\section{Methods}
\subsection{Dataset}

We used the KiTS19 Challenge database \cite{heller2019kits19}. 
It consists of 300 contrast enhanced CT scans, acquired in the pre-operative arterial phase and selected from a cohort of subjects who underwent a partial or a radical nephrectomy between 2010 and 2018 at the university of Minnesota. The included volumes are characterized by different in plane resolutions ranging from 0.437 to 1.04 mm, with a slice thickness varying among cases from a minimum of 0.5 mm up to 5.0 mm.

The dataset provides also for each included case the ground truth mask of both tumor tissue and healthy kidney tissue. Ground truth labels have been manually created by a pool of medical students under the supervision of an expert clinician, by using only image axial projections. A detailed description of the ground truth segmentation strategy is described in \cite{heller2019kits19}.

\subsection{Preprocessing}
Before training our multi-stage model, data were standardized to take into account differences and reduce heterogeneities between the available scans. A re-slicing operation was first performed bringing all volumes to the same slice thickness of 3 mm. This value was chosen as a compromise between the most common axial resolutions observed in the dataset. 

Secondly, the training data values were bracketed in the Hounsfield Unit (HU) range of -30 HU and 300 HU. Hence it was possible to easily remove from the segmentation the fat regions surrounding the kidney, that are characterized by values smaller than -30 HU \cite{heller2019kits19}. On the other hand, the higher threshold was chosen to highlight the high intensity structures and borders that generally characterize the cancerous tissue. After that, data were standardized to a zero mean and unit variance distribution of the pixel values.

\begin{figure*}[htbp!]
      \centering
          \includegraphics[width=0.95\textwidth]{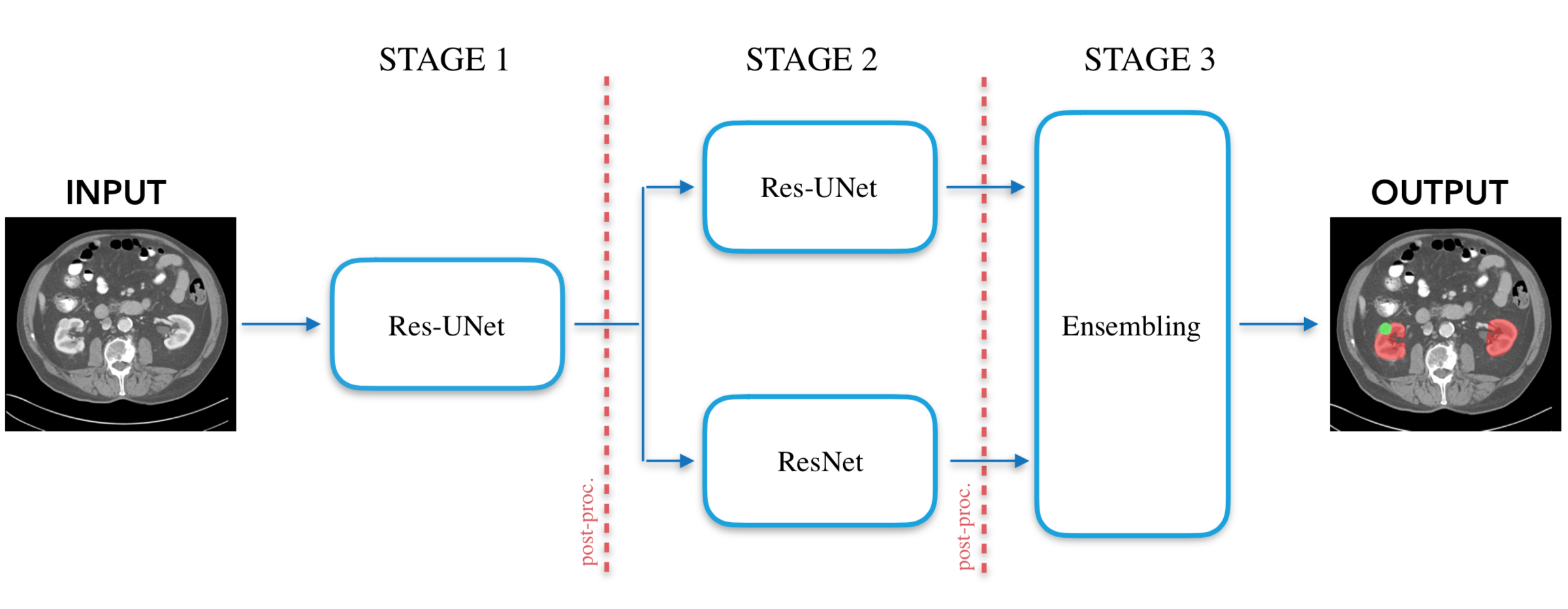}
      \caption{Block diagram of the proposed segmentation model.}
      \label{fig:figure1}
\end{figure*}
\subsection{Multi-stage Deep learning approach}
The proposed model is characterized by three different stages, as summarized in Figure \ref{fig:figure1}. The first stage aims to roughly segment the region of interest (ROI) where to focalize the subsequent analysis, in this case the kidney region. In the second stage, the segmentation of the kidneys and cancerous tissue, is carried out by two different neural networks, which work on the image sub-portions extracted thanks to the use of the approximate kidney predictions from stage one. The results are finally combined in the last stage, where the final segmentation is obtained by using an ensembling operation. In the following, we provide details about the implementation of each stage.

\begin{figure*}[htbp!]
      \centering
          \includegraphics[width=0.95\textwidth]{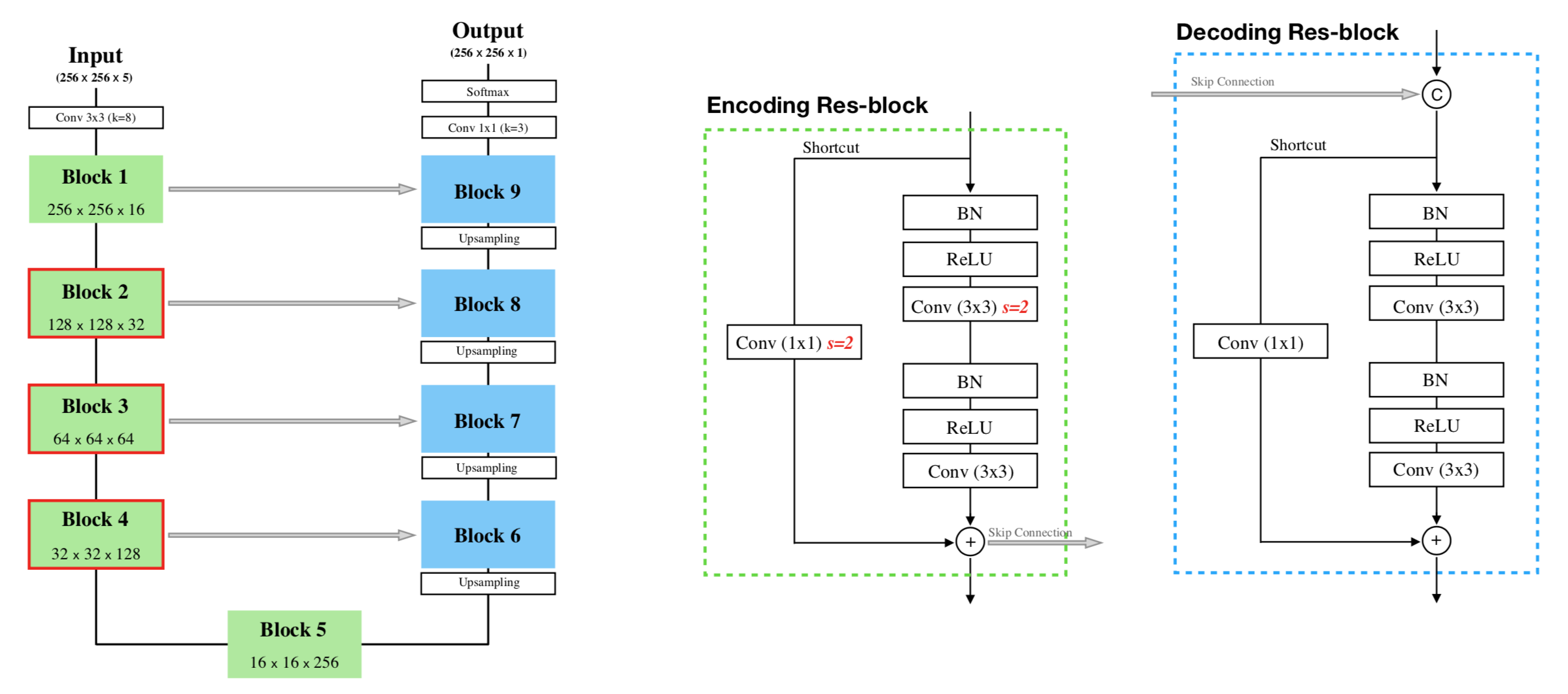}
      \caption{Res-UNet architecture. On the picture left side, the residual block configurations. Marked in red the residual blocks where stride 2 convolutions are used. Upsampling layers precede the decoding res-blocks. }
      \label{fig:figure2}
\end{figure*}

\subsubsection{Stage 1:}
The initial kidney identification was performed using a Residual-UNet (or Res-UNet) designed with a standard encoder-decoder structure of a UNet \cite{ronneberger2015u}. Particularly, we built it with four encoding levels, as many to recreate the final prediction mask. For each level we used a pre-activated residual block \cite{he2016identity} characterized by two convolutional layers and a shortcut connection to add the block input to its output, with the aim of speeding up the model convergence. In all the shortcut connections a convolutional layer was also added to fix some dimension differences occurring in the feature map processing. In all the residual blocks of both encoding and decoding paths, a 1$\times$1 convolution was added to adjust the number of channels and make the addition operation possible, without any dimension mismatches. Moreover, in residual blocks two to four (highlighted in red in Figure \ref{fig:figure2}) we used a 1$\times$1 convolution with stride two to cope with the size reduction passing from a level to another. In all cases a linear activation function was adopted, in order to make the connection purpose as close as possible to the original one, proposed in \cite{he2016identity}.

To upsample the data and restore the original dimension a transposed convolution with stride two was applied before every residual block in the expanding path of the network. Skip connections were finally included to concatenate low level, but with higher resolution, feature maps from the encoder to the high level features in the decoding part. A detailed description can be seen in Figure \ref{fig:figure2}.

As our purpose in the first stage was to roughly segment kidneys with cancerous tissue we tried to enhance the perception of global information rather than local ones by feeding the Res-UNet with subsampled versions of the original images, halved at 256$\times$256 pixels in the x-y plane. Moreover, instead of a full 3D, we employed a 2.5D approach. 

Using 2.5D inputs generally means to provide the network 2D slices coming from coronal, axial and sagittal planes \cite{roth2014new, wolterink2016automatic, setio2016pulmonary, roth2016deep}. In this case it consisted in passing to the model a series of adjacent axial slices stacked together along the channel dimension and make the network able to predict a mask corresponding only to the central slice. 

The final prediction was carried out with a softmax function to assess whether pixels belonged to kidneys, tumor or background. Even if a three-labelled mask was produced, we merged the last two (kidneys and tumor labels) in a single meta-class and we used it in the subsequent stage to localize kidney region and extract two ROIs, i.e. one for each kidney, if present.
\begin{figure*}[htbp!]
      \centering
          \includegraphics[width=0.99\textwidth]{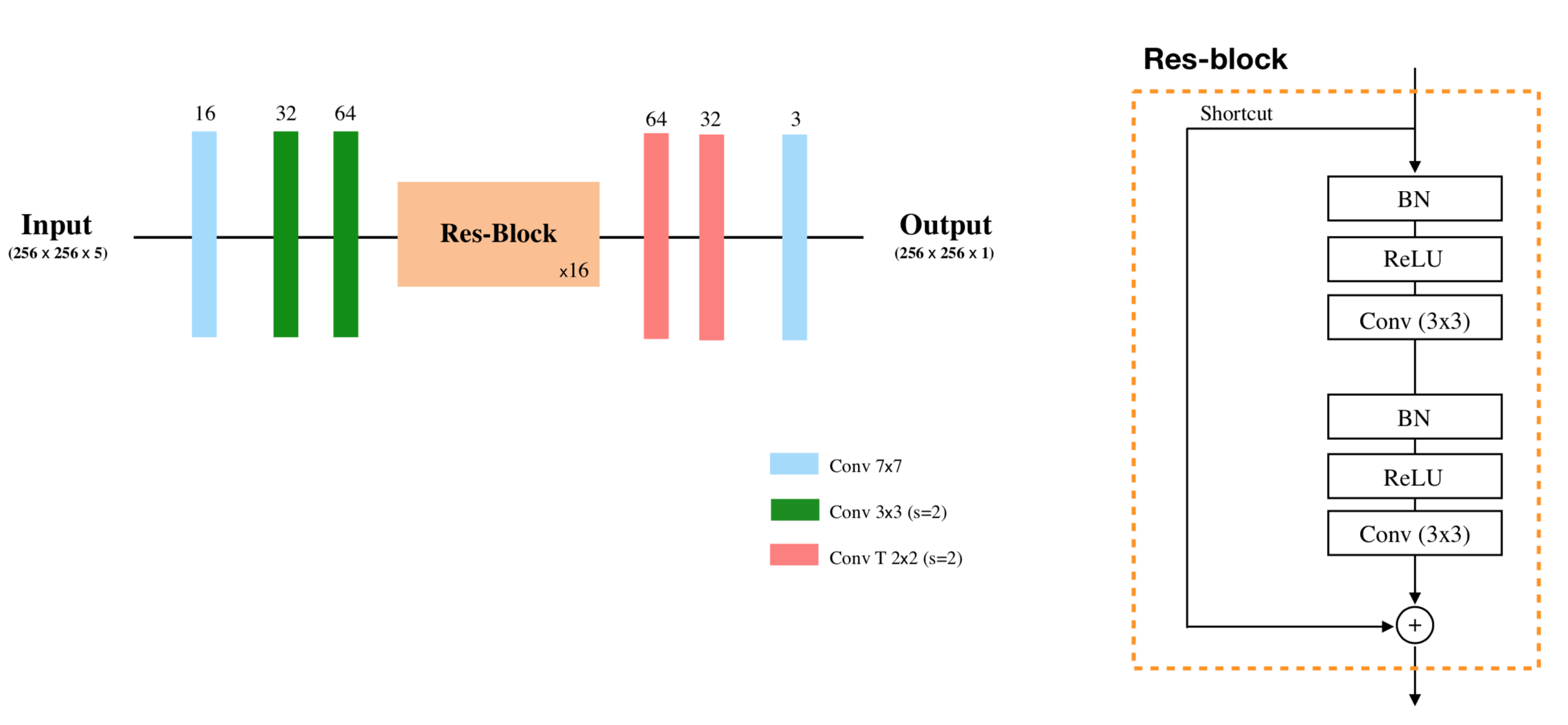}
      \caption{Res-Net architecture. On top of each layer the number of kernels is specified.}
      \label{fig:figure3}
\end{figure*}

\subsubsection{Stage 2:}
As presented above, this stage was used to perform the actual segmentation of both kidney and cancerous tissue. We employed two different convolutional neural network (CNN) architectures to this aim: the first one is identical to the Res-UNet of stage one, while the second network resumes a model proposed in \cite{johnson2016perceptual} 
and extended in another work \cite{van2019automatic} 
to solve a different segmentation task. (i.e. the SegTHOR challenge).

Compared to \cite{van2019automatic} we adopted the same network architecture (see Figure 3) but we simply modified it in order to work with 2.5D input images, that are bigger in size, and directly predict three classes with a softmax classifier. Dropout layer (p=0.5) and batch normalization with pre-activation were used throughout all the residual blocks.

The aforementioned networks were both trained trying to minimize a weighted categorical cross-entropy, with image sub-portions at the original full resolution along x-y and an interpolated slice thickness at 3 mm. 

The ROIs to segment were extracted creating a bounding box to circumscribe the union of the kidney plus the tumor (or just the kidneys) by using the (x, y, z) coordinates from the first stage segmentations. After that, the bounding box was symmetrically expanded to reach the final size of 256$\times$256 pixels. We experimentally verified that such dimension allowed to fully include the structures of interest inside the ROIs, even in case of extended tissue anomalies. Moreover, such input configuration allowed to reuse the first stage network configuration avoiding an interpolation process on the extracted images.

\subsubsection{Stage 3:}
The proposed model ends with an ensembling stage, which combines prediction masks coming from the two networks of stage two. A higher segmentation accuracy is indeed demonstrated by combining more models with respect to the use of the single ones \cite{lyksborg2015ensemble, kamnitsas2017ensembles}.

\subsection{Post-Processing}
A simple post-processing operation was carried out at the end of both the first and second stages (red dashed lines in Figure \ref{fig:figure1}). It mainly consisted in a labelling operation performed on the predicted segmentations, to identify and subsequently remove additional disconnected structures that could occur around kidneys or in other image positions. Such false positive structures were filtered away by counting the number of pixels belonging to every labeled object detected, leaving only the ones composed by more than 5000 pixels. We expected to find the kidneys with or without cancerous tissue as the largest connected structures after the two segmentation stages.

\subsection{Evaluation metric}
As proposed by the challenge organizers, the output image segmentation quality was assessed with the Dice score index (Eq. \ref{eq:dice}) computed on tumor and kidneys, considered as a single entity and on tumor as a standalone object. Both structures segmented with our method ($S_{DL}$) were compared with the ground truth segmentations ($S_{GT}$) provided.

\begin{equation}
  Dice(S_{GT}, S_{DL}) =  \frac{2 \cdot (S_{GT} \cap S_{DL})}{|S_{GT} + S_{DL}|}
 \label{eq:dice}
\end{equation}

\section{Experimental settings and results}
From 210 patients we received in the first round, we used 190 cases to train our model and the remaining 20 to validate it. The final 90 cases were provided by the organizers without the ground truth masks and were used to rank the methods proposed in the context of the challenge according to the average score obtained on the 90 test cases, combining the tumor and the kidney dice indexes.

\subsection{Training settings}
We designed our networks using Tensorflow \cite{abadi2016tensorflow} 
and trained them from scratch on single NVIDIA GTX 1080 of 11 GB, minimizing in all cases a weighted categorical cross-entropy, and speeding up the process using an Adam optimizer. An L2 regularization on kernel weights with 0.1 scale factor was also used.

250 epochs were fixed as upper bound limit for training each single network, but in all the conducted trials, the best model was always reached around 170 $\pm$ 30 epochs.

As anticipated above, all networks were trained with 2.5D input images resulting from the concatenation of two slices above and two slices below the current axial slice to segment.

In every training iteration, a balanced batch size of 32 samples was used. During stage one, the batch cases were balanced according to three image group types (roughly 33\% each) characterized as follows. \textit{Group B}: images in which neither kidneys nor tumors appear, \textit{group K}: images with healthy kidney portions, \textit{group KT}: images with kidney and tumor tissue. For stage two, 50\% of cases belonged to \textit{group K}, while the remaining 50\% was filled with cases from \textit{group KT}.

In order to prevent overfitting on the tumor segmentation, different data augmentation strategies such as axial rotation (angle $\in [-30^{\circ}, 30^{\circ}]$), horizontal flip, and central crop plus zoom were also employed on-the-fly on KT cases only, as they appear in the training dataset less frequently and with higher variability than the K cases. 

The use of diverse combinations of initializations, training settings and network architectures generally help subsequent ensembling operations \cite{lyksborg2015ensemble, kamnitsas2017ensembles}. For this reason, we carried out different training configurations (especially for stage 2) in order to reduce the generalization error of the final prediction. Table \ref{tab:table1} synthesizes different training configurations adopted for the networks employed.

\begin{table}
\caption{Training configurations and data augmentation strategies employed for each network. Data augmentation operations refer only to input images presenting tumors (group KT). T.N. stands for Truncated Normal weights initialization.}
\makebox[\textwidth][c]{
  \begin{tabular}{lSSSSSS}
    \toprule
    \multirow{3}{*}{ } &
      \multicolumn{3}{c}{\bf{Data Augmentation}} &
      \multicolumn{3}{c}{\bf{Training Settings}} \\
      \cmidrule(lr){2-4}\cmidrule(lr){5-7}
      & {Rotation} & {H-Flip} & {Central crop} & { Weights} & {Loss weights} & {Learning} \\
      & {$(p=1.0)$} & {$(p=0.5)$} & {$(p=0.66)$} & {Init} & {[B, K, KT]} & {rate } \\
      \midrule
    Res-Unet1 & $\bullet$ &   &   & {T.N. ($std\  0.1$)} & {[0.3 \ 1.0 \ 3.0]}  & e-4 \\
    Res-Unet2 & $\bullet$ & $\bullet$ &   & {T.N. ($std\  0.1$)} & {[0.3 \ 1.0 \ 3.0]} & e-4 \\
    Res-Net & $\bullet$ & $\bullet$ & $\bullet$ & {He uniform} & {[0.2 \ 0.25 \ 0.55]} & e-3\\
    \bottomrule
  \end{tabular}
  }
  \label{tab:table1}
\end{table}

\subsection{Evaluation results}
With the above configurations we obtained the results expressed in Table \ref{tab:table2}, on the 20 cases chosen randomly as validation set. We finally used the trained ensemble model to perform the prediction on the 90 test cases. We obtained a dice score of 0.96 and 0.74 for kidneys and tumor respectively. Segmentation image examples follow.

\begin{table}
\caption{Segmentation results for each stage designed, considering kidneys plus tumor (first column) and tumor alone (second column) on the validation set.}
\makebox[\textwidth][c]{
  \begin{tabular}{lSS}
    \toprule
    \multirow{3}{*}{ } &
     {\bf{Dice kidneys\  \ \ }} & {\bf{\ \ Dice tumor}} \\
      \midrule
    Res-Unet1 & {$0.96 \pm 0.02$}\  & \ {$0.52 \pm 0.32$}    \\
    Res-Unet2 & {$0.97 \pm0.02$}\  & \ {$0.70 \pm 0.28$}  \\
    Res-Net & {$0.97 \pm 0.02$}\  & \ {$0.72 \pm 0.26$} \\
    \midrule
    Ensembling & {$\bf{0.98 \pm 0.01}$}\  & \ {$\bf{0.73 \pm 0.25}$}\\
    \bottomrule
  \end{tabular}
  }
  \label{tab:table2}
\end{table}

\begin{figure*}[htbp!]
      \centering
          \includegraphics[width=1.0\textwidth]{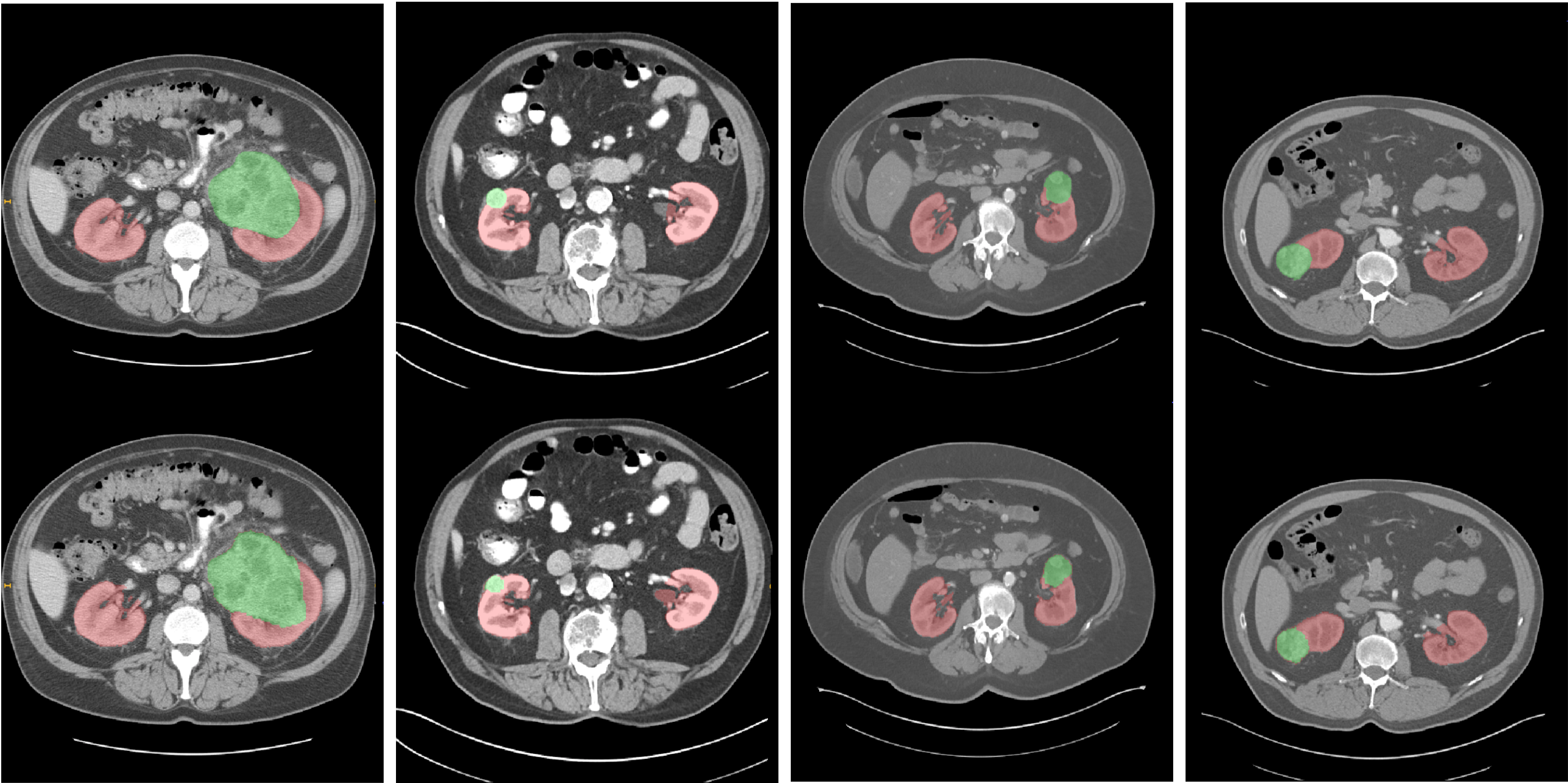}
      \caption{Segmentation examples from four patients. On the top row the ground truth labels are reported (red for kidneys, green for tumors). The proposed model predictions are on the second row. Best viewed in color.}
      \label{fig:figure4} 
\end{figure*}
\section{Discussion and conclusion}
We presented an automatic method for semantic segmentation of kidneys and kidney cancerous tissue from contrastographic CT acquisitions. 

As shown in Table 2, the use of a multi-step segmentation algorithm allowed us to obtain better results than using only a single segmenting step. Moreover, the ensembling operation of different CNN predictions proved to give slightly better results, compared to the individual network performances. 

On the other hand, the use of an ensembling training strategy combined with a multi-stage segmentation process led to an increase in training time, due to the need to try more networks with different configuration settings. A single network took indeed around three days of training with our hardware configuration. 

The use of a relatively large batch size allowed to better exploit batch normalization properties compared to smaller batches previously tried (8 or 16). Among the evaluations, the results obtained with a batch size of 32 samples were better. However, this choice penalized the use of state-of-the-art neural network architectures like Tiramisu \cite{jegou2017one} 
or DeepLab \cite{chen2017deeplab}, 
which couldn't be trained with such input sizes, essentially due to memory constraints.

Finally, using 2.5D input tensor helped to provide some kind of volumetric information to the network resulting in a better segmentation level compared to simple 2D inputs.

Considering the final performance obtained from the 90 test patients, we can assert that the overall segmentation results, especially concerning the tumor identification remain promising, but are quite low as a consequence of false positive cases that were sometimes detected on kidney cysts, or false negative cases where the tumor lesion was not east to identify on the CT. For future improvements we plan to design a new training strategy, in which more of these specific cases could be passed to the model, in order to differentiate them from cases where cancerous tissues actually occur. Furthermore, the we would like to work on even more focused input data, by maybe adding an additional stage in order to focus on the analysis of even more detailed local features.

\section*{Acknowledgments}
This work is financed by FEDER funds through "Programme op\'erationnel regional FEDER-FSE Pays de la Loire 2014-2020" n$^{\circ}$ PL0015129 (EPICURE).

%
%
%






%
%

\bibliographystyle{splncs}
\bibliography{kits19_biblio}

\end{document}